\documentclass[epj]{svjour}
\usepackage{graphics}
\begin{document}
\title{Sudden Critical Current Drops Induced in S/F Structures}
%
\author{
%
Edgar J. Pati\~{n}o\inst{1,2}, Chris Bell\inst{1}, \and Mark G.
Blamire\inst{1}
}                     
\offprints{Edgar J. Pati\~{n}o}          
\institute{Materials Science Department, IRC in Superconductivity
and IRC in Nanotechnology, University of Cambridge, Cambridge,
United Kingdom \and Departamento de F\'isica, Grupo de Física de
la Materia Condensada, Universidad de los Andes, Bogot\'a,
Colombia.}


\date{Received: date / Revised version: date}
%
\abstract{In the search for new physical properties of S/F
structures, we have found that the superconductor critical current
can be controlled by the domain state of the neighboring
ferromagnet. The superconductor is a thin wire of thickness $d_{s}
\approx 2\xi_{S}$. Nb/Co and Nb/Py (Permalloy Ni$_{80}$Fe$_{20}$)
bilayer structures were grown with a significant magnetic
anisotropy. Critical current measurements of Nb/Co structures with
ferromagnet thickness $d_{F}>30$~nm show sudden drops in two very
defined steps when the measurements are made along the hard axes
direction (i.e. current track parallel to hard anisotropy axes
direction). These drops disappear when they are made along the easy
axis direction or when the ferromagnet thickness is below 30~nm. The
drops are accompanied by vortex flux flow. In addition
magnetorestistance measurements close to $T_{C}$ show a sharp
increase near saturation fields of the ferromagnet. Similar results
are reproduced in Nb/Py bilayer structure with the ferromagnet
thickness $d_{F}\sim50$~nm along the easy anisotropy axes. These
results are explained as being due to spontaneous vortex formation
and flow induced by Bloch domain walls of the ferromagnet
underneath. We argue these Bloch domain walls produce a 2D
vortex-antivortex lattice structure.
}
\maketitle
\section{Introduction}
\label{intro}
The proximity effect in superconductor (S) ferromagnet (F) bilayer
structures has been subject of intensive research in the past ten
years. This has been specially due to the effect exchange field
has on Cooper pairs generating the so called Fulde Ferrel Larkin
Ovchinnilov (FFLO) state. In this state the real part of the order
parameter penetrates the ferromagnet and oscillates in proximity
to the interface leading to the theoretically predicted
oscillations of $T_{C}$ as a function of the ferromagnet thickness
\cite{radovic} and the experimentally verified $\pi$ phase shift
in S/F multilayers \cite{ryazanov0,jiang}.

In addition to these investigations, other experiments carried out
by Kinsey \emph{et al.} \cite{Kinsey} studied the effect that
domain wall formation has on superconductivity. These were carried
out on Nb/Co bilayers, with 15-65 nm of Nb on 54$\pm$9 nm of Co,
by critical current measurements of patterned films using
photolithography. The geometry of the mask used for this process
included the voltage contacts perpendicular to the current track,
that leaded to stray fields on the current track. In these earlier
studies an increase in the critical current measurements of the
patterned wire was found by applying a magnetic field of the order
of the coercive field of the ferromagnet. This effect was
attributed to a reduction of the average exchange field sampled by
the Cooper pairs within the domain wall giving rise to an
enhancement of superconductivity so called domain wall
superconductivity (DWS)\cite{buzdin}. However, in these
experiments the effect of stray fields, from the mask geometry
used, in the measurements could not be completely eliminated.
Later a similar effect was reported in Nb/CuNi trilayers by
Rusanov \emph{et al.} \cite{Rusanov} again in this case the stray
fields from the mask geometry were present (i.e. magnetic voltage
contacts perpendicular to the current track). In an experiment, by
the same authors \cite{Rusanov2}, on Nb/Py (Permalloy
Ni$_{80}$Fe$_{20}$),with 21 nm of Nb and 20 nm of Py, suggests
evidence of DWS. In this experiment Rusanov \emph{et al.} found
enhancement of superconductivity at coercive fields of the
ferromagnet, by measuring magnetoresistance at the transition
region between normal and superconducting state. Dips in magneto
resistance measurements at coercive fields were attributed to be
evidence of this effect. Nevertheless this effect was only found
for large samples ($0.5$~mm~$\times$~4~mm) and absent in small
samples (1.5~$\mu$m~$\times$~20~$\mu$m), where the authors argued
that no stable domains form in small samples. In other experiments
on Nb/CuNi bilayers performed by Ryazanov \emph{et al.}
\cite{ryazanov}, they found magnetoresistive peaks in differential
resistance measurements at coercive field of the ferromagnet. They
explained this observations by spontaneous vortex formation and
flow from stray fields from the domain walls of the ferromagnet. A
theoretical work has been published by Burmistrov \emph{et al.}
\cite{burmistrov} investigating the effect of Bloch domain walls
on the current distribution in the superconductor. They also
determine a lower critical value of domain wall magnetization
above which vortex formation is favorable.
Based on this idea experiments, performed by Steiner \emph{et al.}
\cite{Steiner} on Fe/Nb and Co/Nb bilayers, found a reduction of
$T_{C}\approx$~0.5$\%$ as a function of magnetic state of the
ferromagnet. Additionally Bell \emph{et al.} \cite{Bell} have
found a double peak structure in magnetoresistance measurements of
MoGe/GdNi bilayered structures as a function of the magnetic state
of the ferromagnet attributed to vortex formation and flux flow.


The aim of the present work is to further investigate the
influence of N\'{e}el and Bloch domain walls in the
superconducting pairing and vortex generation in S/F hybrids. This
investigation differs from previous ones in that it considers a
wide range of ferromagnet's thickness and eliminates stray fields
from the mask geometry.
We also used Co and Py as ferromagnets where the domain wall
structures is fairly well understood at least at room temperature.
\section{Critical Current and Magneto-Resistance Measurements}
\label{measur}
In our experiment we have grown Nb/Co and Nb/Py bilayers (Nb on top)
on Si (100) substrates using a UHV DC-magnetron sputtering system in
a chamber cooled to $-100^{o}$C using liquid nitrogen. The base
pressure was less than $3 \times 10^{-9}$ mbar. The partial oxygen
base pressure measured using a mass spectrometer showed a value less
than $0.1 \times 10^{-10}$ mbar. The Nb, Co and Py sputtering
targets were 99.95 $\%$ pure, deposited under Ar pressure of 0.5 Pa,
in an in-plane magnetic field H of approximately 400~Oe. Under these
conditions the F layer in the bilayer structures grown had a strong
magnetic anisotropy showing easy and hard axes anisotropy and
coercive fields of about 35 and 16~Oe for the Co and 4.5 and 1.2~Oe
for Py respectively. Several bilayer structures with Nb thickness
$d_{Nb}$ = 25~nm (i.e. of twice the superconducting low temperature
coherence length, $\xi_{S}\approx 12$~nm) and Co thickness,
$d_{Co}$, between 2 and 50~nm were made where $d_{Co}\gg
\xi_{Co}\approx 0.3$~nm, the ferromagnet's coherence length. The
interface RMS roughness, measured using AFM, was about 0.3 nm. The
superconducting transition temperature $T_{C}$ for these Nb/Co
bilayer structures determined by four point resistivity measurements
had an average value of 6.4 $\pm$ 1 K depressed by the proximity
effect, and a small transition width of 100 mK demonstrating the
very high uniformity and quality
of our films. \\
In order to compare the results the study was finalized by replacing
the Co layer with Py with a thickness of 50~nm. For this thickness
all the results obtained for Co and discussed in this paper
reproduce qualitatively the same for Py.

Critical current measurements were made on bilayer structures
patterned into tracks. Usually most experimental configurations for
doing four point measurements of narrow tracks include; the voltage
contacts within the same structure and current tracks with sharp
edges. Due to the presence of the ferromagnet in the S/F
hybrid-structures the free surfaces from, the voltage contacts or
sharp edges, perpendicular to the ferromagnet's magnetization form
free magnetic poles which effectively act as ``external magnets''.
Depending upon the geometry and type of ferromagnet used these
source of stray fields affect the superconductor to a lower or
greater extent.

The current track was patterned using a mask without sharp edges at
the center, where the measurements are made. The shape of the mask
should mean that if there are free poles they contribute very little
stray fields to the measurement section of the track. The voltage is
measured using non magnetic Cu contacts. These prevent the wires
acting as ``external magnets''.
The center of the mask used to make the device is shown in Fig.~\ref
{IV-dep2}.
The geometry of the designed mask for the photolithography consists
of three different current tracks of 2, 5 and 7~$\mu$m wide (drawn
in blue color). This allowed us to take measurements with different
aspect ratios. Only further away from the center the tracks width
gradually increases in size to avoid sharp edges. Non magnetic Cu
contacts (black color), spaced $\approx$7~$\mu$m, were deposited by
sputter deposition in order to measure the voltage. This eliminates
stray fields from the voltage contacts. The Cu contacts also play an
important role in dissipating the heat generated by the Nb track in
the normal state.

Current voltage (I-V) characteristics have been obtained. A plot of
typical I-V characteristics at different temperatures is shown in
Fig.~\ref {IV-dep}. A clear jump from the normal to the
superconducting state determines the value of the critical current
($I_{dp}$). The small voltage onset below $I_{dp}$ is due to vortex
motion, unavoidable close to $T_{C}$.
\begin{figure}
%
%
\resizebox{0.47\textwidth}{!}{%
\includegraphics{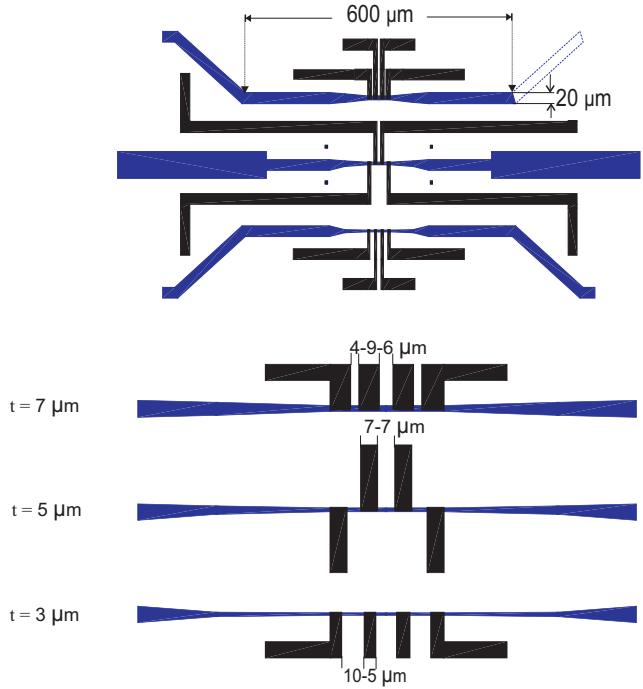}
}
\caption{\label{IV-dep2} Diagram of the mask used in experiment
including dimensions in microns. The figures show an enlargement of
the mask middle section. The current tracks pattern is drawn in blue
and the voltage contacts pattern is drawn in black color. This mask
gives the possibility to pattern tracks of three different widths:
7, 5, and 3 $\mu$m (zoomed view bottom the figure).}
\label{IV-dep2}       
\end{figure}
\begin{figure}
%
%
\resizebox{0.47\textwidth}{!}{%
\includegraphics{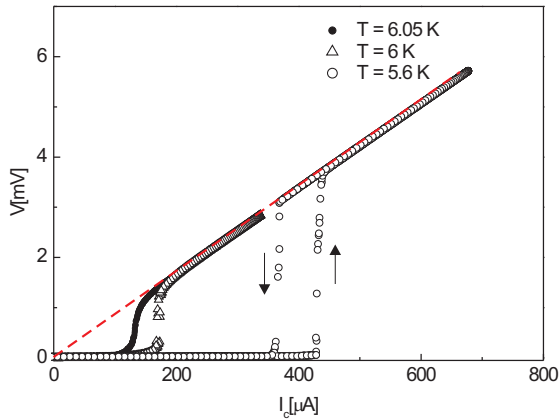}
}
\caption{\label{IV-dep} Current-voltage characteristics, taken at
zero field, shows when the normal state is reached. The largest
value of current below the voltage onset defines the critical
current value. I-V characteristics are shown for three different
temperatures, dashed line indicates the normal state resistance.}
\label{IV-dep}       
\end{figure}

First Nb/Co bilayers were patterned into tracks along the hard axis
direction of the ferromagnet and then in-plane critical current
measurements were made with the field parallel to the track. Between
a Cobalt thickness of 2~nm and 30~nm no appreciable changes in
critical current in Nb/Co bilayer structures were observed. However,
for a ferromagnet thickness $d_{F}\geq$30~nm critical current
measurements exhibit sudden drops at fields of the order of the
coercive and saturation fields.
The dips observed on a 25nm Nb on 42~nm Co, depicted in
Fig.~\ref{dips2542}, show up to a 50 $\%$ reduction from its maximum
value.
\begin{figure}
%
\resizebox{0.50\textwidth}{!}{%
\includegraphics{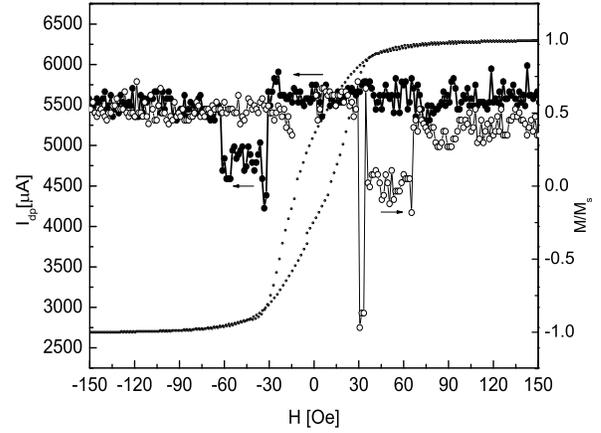}
}
%
%
\caption{\label{dips2542} Critical current vs. in plane magnetic
field of a Nb(25)/Co(42) sample taken at $T=$4.2~K with
I$\parallel$H. Arrows show the field-scan direction. Superimposed is
the corresponding hysteresis loop.}
\label{dips2542}       
\end{figure}
The drops increase in magnitude with temperature and thickness of
the ferromagnet. However when taking the same measurements either
with the field perpendicular to the track or by patterning the track
along the easy axis direction the dips disappeared. This indicates a
strong correlation with the anisotropy axes of the Cobalt.

Finally magnetoresistance measurements were made after stabilizing
the temperature near $T_{C}$ at the transition region just above the
zero resistance. A sudden increase in magnetoresistance as the one
depicted in Fig.~\ref{magneto} for a sample of 25~nm Nb on 50~nm Py
(Nb(25)/Py(50)), is observed at the same field range for each of the
samples where the dips occur.
%
%
Note that in our experiments both; critical current and
magnetoresistance measurements (Fig.~\ref{magneto} and
\ref{dips2542}) the
dips and peaks obtained show two defined steps 
which will be discussed later.

\begin{figure}
%
\resizebox{0.50\textwidth}{!}{%
\includegraphics{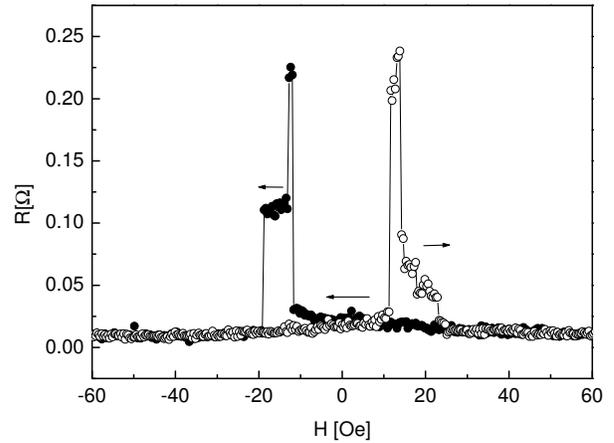}
}
%
%
\caption{\label{magneto} Resistance as a function of applied field
of a patterned sample Nb(25)/Py(50), with I$\parallel$H along the
easy anisotropy axes, measured at 6.06 K where $T_{C}\approx6.11$~K.
Arrows show the field scan direction.}
\label{magneto}       
\end{figure}

In order to understand these results the I-V characteristics were
analyzed close to $T_{C}$ at the dip region. As shown in the inset
in Fig.~\ref {fluxindips} when the applied field is increased from
-25 Oe towards -30 Oe where the dip in critical current takes place
a voltage onset below the normal critical \cite{note} current value
is observed indicating that vortex motion is involved. Because the
measurements are made in a Lorentz force free configuration (current
parallel to the field) this vortex motion indicates the presence of
vortices perpendicular to the plane which causes this voltage
onset.These are shown in the inset of Fig.~\ref{fluxindips}.

%
%
%
\begin{figure}
%
\resizebox{0.45\textwidth}{!}{%
\includegraphics{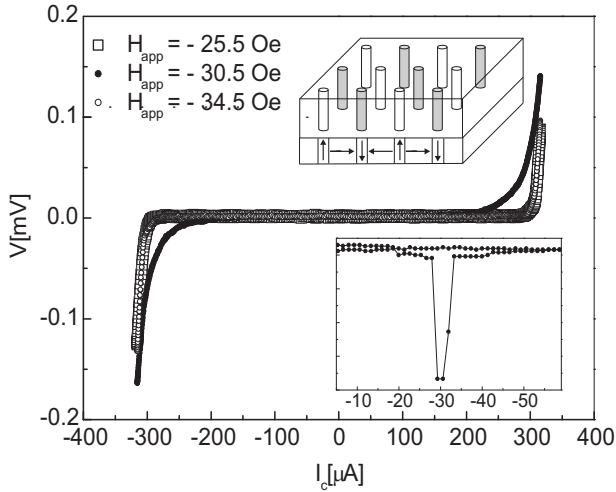}
}
%
%
\caption{\label{fluxindips} IV characteristic of Nb(25)Co(50) sample
taken close to the transition $\approx$~0.9$T_{C}$ for different
values of field with I$\parallel$H. Motion of vortices is indicated
by voltage onset below the normal critical current value. Inset
shows vortex-antivortex array distinguished by clear and dark color
induced in the superconductor by the Bloch domain structure of the
ferromagnet underneath.}
\label{fluxindips}       
\end{figure}

\section{Analysis of Results}
\label{analysis}
In this letter we present a mechanism by which the
experimental observations can be explained. As pointed out
previously the magnitude of the dips depends on the ferromagnet's
thickness and are only observable beyond a thickness of the
ferromagnet $d_{F}$ $>$ 30~nm. This is of the order of thickness
where domain walls of the ferromagnet, Co or Py, are thought to
gradually mutate from being in plane i.e. N\'{e}el domain walls
towards the out of plane direction Bloch domain walls.
At the range of film thickness; between 35 to 100 nm the domain
structure is formed by cross tie domain walls.

For Co films with a thickness of 50~nm cross-tie domain walls have
been experimentally confirmed using magnetic force microscopy \cite
{crosstieco50}. In this case Bloch lines are located periodically
with alternating polarity, i.e. pointing outwards or inwards
producing high stray fields in this direction, resembling a
checkerboard of antiparallel stray fields.

Similar observations for Py films in the domain structure, were
found using the Kerr effect, for thicknesses between 30 and 200~nm
\cite {crosstiepy}. The out of plane stray field above a Bloch line
can be roughly estimated for 50~nm Co films based on the
observations reported in \cite {crosstieco50}. Here each Bloch line
has an approximate width $\delta\approx 0.15 - 0.3$~$\mu$m with
magnetic fields $>$~2800~Oe and a magnetic flux between 3-13
$\Phi_{0}$, where $\Phi_{0}$ is the flux quantum. Moreover
considering the number of Bloch lines, pointing in the same
direction, divided by the interface area, an out of plane effective
field of $H_{eff}^{BDW} \approx \pm 48$~Oe is calculated (this is
the maximum value of stray field used later as $B_{o}$ parameter in
Eq. \ref{HeffBDWmiddle}). Since the lower critical field for a thin
Nb film $H_{c1}(0)$ in the out of the plane direction is about
10-20~Oe, vortex formation perpendicular to the film is expected.\\
Furthermore due to the fact that Bloch lines are antiparallel to one
another a 2D vortex-antivortex array pinned by the domain
walls must form as shown in the inset Fig.~\ref{fluxindips}.\\
Note also that as the film gets thicker the domains reduce in size,
increasing the number of domain walls and Bloch lines that can fit
in the same interface area \cite{domsize} thus augmenting
$H_{eff}^{BDW}$ on the superconductor. These leads to a lower
critical current as observed in our experiments after increasing the
ferromagnet's thickness.

\begin{figure}
\resizebox{0.45\textwidth}{!}{%
\includegraphics{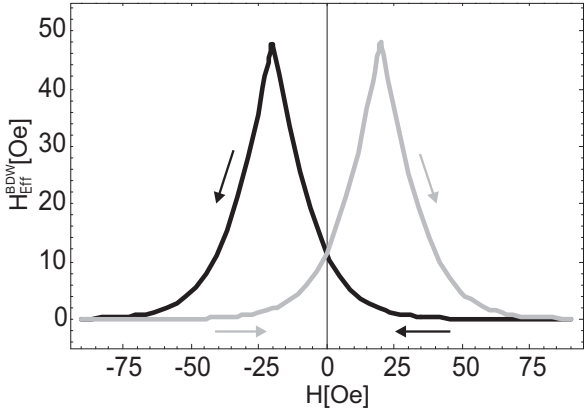}
} \caption{\label{ictheory1}
Stray field from Bloch domain walls is modelled using
expression~\ref{HeffBDWmiddle}. Here $H$ represents the external
field applied within the plane of the film while the effective field
$H_{eff}^{BDW}$ develops along the out of plane direction of the
film as a double peak structure as a function of applied field.}
%
\end{figure}

To have a better understanding of the observed experimental effects
the following simple model is employed. As the ferromagnet,
initially in the remanent state, lowers its magnetization as result
of an in-plane external field $H$, the number of Bloch domain walls
$N$ increases augmenting the effective field $H_{eff}^{BDW}$ (out of
plane). This field reaches its maximum value $B_{o}$ at coercive
field $H_{c}$ and begins to decrease for $H>H_{c}$. This behavior
has been modeled using the following expression:
\begin{equation}\label{HeffBDWmiddle}
    H_{eff}^{BDW}=B_{o}\left(1-\left|\frac{M(H)}{M_{S}}\right|\right)
\end{equation}
where $M_{S}$ is the saturation magnetization.\\
Using the value of $H_{c}=20$~Oe for a Co film and a value of
$B_{o}=48$~Oe (estimated in the previous paragraph) for
Eq.~\ref{HeffBDWmiddle}, a double peak structure of $H_{eff}^{BDW}$
is obtained as shown in Fig.~\ref{ictheory1}. This explains the
results since when the ferromagnet is in a single domain state the
patterned track is in a Lorentz force free configuration (i.e.
current parallel to the field). However when the ferromagnet breaks
into domains $H_{eff}^{BDW}$ develops and the patterned track shifts
to a Lorentz force configuration where the current is now
perpendicular to $H_{eff}^{BDW}$ resulting in a lower critical
current measurement.
\\
%
%
\\

\subsection{Discussion}
\label{discussion}

The absence of any anomalous observations between Cobalt thicknesses
2~nm$<d_{Co} < 30$~nm, permits us to conclude that previous
observations \cite{Kinsey} were probably due to either stray fields
from the device geometry or dipole stray fields.

Regarding the observations in magnetoresistance measurements made
on Nb/Py bilayers \cite{Rusanov2}, the observation of dips is
accompanied by peaks in magnetoresistance measurements. These
imply the presence of stray fields and thus should be considered
in the interpretation of the results. Recently the same author
reported additional results on Nb/Py bilayers \cite{rusanov3},
where two different thicknesses of Py were used; 20 and 60 nm.
Here small dips and large peaks in magnetoresistance were
respectively found. The experiment description indicates that the
device geometry includes magnetic voltage contacts; thus is likely
that these dips are due to stray fields. However the large peaks
for a Py thickness of 60 nm are explained by the authors in
\cite{rusanov3} as the result of Bloch domain wall stray fields.
This result is qualitatively similar to the one shown in
Fig.~\ref{magneto} and previously described on \cite{thesis} by
the present author. The main difference to our results is the two
step structure observed in figures~\ref{magneto}
and~\ref{dips2542}.

Although the previous suggested mechanism in the research
investigations \cite{Kinsey,Rusanov2,rusanov3} (the enhancement of
superconductivity at the domain walls i.e. DWS) is probably
correct, we believe that an unambiguous
experimental evidence of DWS has not yet been successfully produced.\\

\section{Conclusions}
\label{conclusions}

In summary the dips between coercive field and saturation come as a
result of votex-antivortex 2D lattice formation (perpendicular to
plane) which reduces the critical current. These are the result of
Bloch lines formed with strong antiparallel stray fields at the
domain walls. Increasing the field further and fully magnetizing the
sample (where Bloch domain walls are absent) results in a re-entrant
behavior to superconductivity as seen in magnetoresistance
measurements.
Similar properties of so called field-induced superconductivity
(FIS) have only been realized by magnetic dots array deposited on
top of a superconducting Pb film \cite{lange} and in 50~nm Niobium
films deposited on the single crystal ferrimagnet
BaFe$_{12}$O$_{19}$. Both experiments were performed with a
magnetization along the perpendicular direction and with an
insulator between the superconductor and ferromagnet \cite{yang}.

Our inference of Bloch domain induced 2D vortex-antivortex array in
the superconductor makes it possible to experimentally investigate
2D vortex crystal lattice without the need of superconductors with
unconventional pairing symmetry or $\pi$- Josephson junctions
configurations \cite{hilgenkamp}.
\\
In both critical and magnetoresistance measurements two sharp steps
in the region where dips and peaks are observed indicate an abrupt
reconfiguration of the 2D vortex-antivortex array. These two well
defined vortex arrangements deserve further investigation, for
example using scanning tunnelling microscopy (STM) with resolution
on the~nm scale to visualize the vortex dynamics as this could be of
great interest for both applied and fundamental research.

Finally, by confining superconductivity to nano dimensions $\leq$
2$\xi_{S}$ together with the effect from Bloch domain wall
formations makes these 
structures very sensitive to small magnetic fields. This opens up
potential applications such as small field detectors and switching
devices.


\begin{thebibliography}{99}
%
\bibitem{radovic} Z. Radovi\'{c}, M. Ledvij, L. Dobrosavljevi\'{c}-Gruji\'{c}, A.I.
Buzdin, and J. R. Clem, Phys. Rev. B \textbf{44}, 759 (1991).
\bibitem{ryazanov0} V. V. Ryazanov, V. A. Oboznov, A. Y. Rusanov, A. V. Veretennikov, A.
A. Golubov, and J. Aarts, Phys. Rev. Lett. \textbf{86}, 2427 (2001).
\bibitem{jiang} J. S. Jiang, D. Davidovi\'{c}, Daniel H. Reich, and C. L. Chien, Phys. Rev. Lett.
\textbf{74}, 314 (1995).
\bibitem{Kinsey} R. J. Kinsey, G. Burnell and M. G. Blamire, IEEE Trans. Appl.
Supercond., \textbf{11}, 904, (2001).
\bibitem{buzdin} M. Houset and A. I. Buzdin, Phys. Rev. B \textbf{74}, 214507, (2006).
\bibitem{Rusanov} A. Rusanov, M. Hesselberth, S. Habraken and J. Aarts, Physica
C, 404, Issue 1-4, 322 (2004).
\bibitem{Rusanov2} A. Yu. Rusanov, M. Hesselberth, J. Aarts and A. I. Buzdin, Phys. Rev.
Lett. \textbf{93}, 57002 (2004).
\bibitem{ryazanov} V. V. Ryazanov, V. A. Oboznov, A.S. Prokof'ev, and S. V. Dubonos,
JETP Lett. \textbf{77}, 39 (2003).
\bibitem{burmistrov} I. S. Burmistrov and N.M. Chtchelkatchev, Phys. Rev. B, \textbf{72}, 144520 (2005).
\bibitem{Steiner} R. Steiner and P. Ziemann, Phys. Rev. B, \textbf{74}, 94504 (2006).
\bibitem{Bell} C. Bell, S. Tursucu and J. Aarts, Phys. Rev. B, \textbf{74}, 214520 (2006).
\bibitem{note} In most of our meassurements well bellow $T_{C}$ we did not observed flux
flow at the dip region but an abrupt reduction of the critical
current value. However as shown in Fig.~\ref {fluxindips} close to
$T_{C}$ the pinning barrier is easily overcomed by thermal
activation.
\bibitem{crosstieco50} M. L\"{o}hndorf, A. Wadas, H. A. M. van den Berg, and R. Wiesendanger, Appl. Phys.
Lett., \textbf{68}, 25, (1996).
\bibitem{crosstiepy} S. Methfessel, S. Middelhoek, H. Thomas, IBM
Journal, 96, (1959).
\bibitem{domsize} A. Stankiewiczyz, S. J. Robinsonz, G. A. Gehringzx and V. V. Tarasenkoy,
J. Phys: Condens. Matter \textbf{9},1019–1030, (1997).
\bibitem{rusanov3}A.Yu. Rusanov, T.E. Golikova, and S.V. Egorov.  JETP Lett. \textbf{87}, 175 (2008)
%
\bibitem{thesis} E. J. Pati\~{n}o, ``Study of The Influence of Domain Walls in The Superconductor/Ferromagnet
Proximity Effect". Ph.D Thesis [Non-Published], Cambridge
University, (2005).
%
%
\bibitem{lange} M. Lange, M. J. Van Bael, Y. Bruynseraede, and V. V.
Moshchalkov, Phys. Rev. Lett. \textbf{90}, 197006 (2003).
\bibitem{yang} Z. Yang, M. Lange, A. Volodin, R. Szymczak and V. V. Moshchalkov, Nature Mat. \textbf{3}, 793, (2004).
\bibitem{hilgenkamp} H. Hilgenkamp, Ariando, Henk-Jan H. Smilde, Dave H. A. Blank, G. Rijnders, H. Rogalla, J. R. Kirtley and C. C. Tsuei,
Nature \textbf{422}, 50 (2003).
%
\end{thebibliography}
\end{document}